\documentclass[twocolumn,floatfix]{revtex4}
\usepackage{epsf,graphicx}
\usepackage{amsmath,amssymb,amsfonts}
\newcommand{\vecbm}[1]{\mbox{\boldmath#1}}

\begin{document}
\title{Nuclear statistics, microcanonical or canonical ?\\ The physicists vs. the chemists approach.} \today
\author{D.H.E. Gross}
\affiliation{Hahn-Meitner Institute and Freie Universit{\"a}t Berlin,\\
Fachbereich Physik.\\ Glienickerstr. 100\\ 14109 Berlin, Germany}
\email{gross@hmi.de} \homepage{http://www.hmi.de/people/gross/ }
\begin{abstract} Nuclei are {\em small and inhomogeneous}.
Multi-fragmented nuclei are even more inhomogeneous and the fragments even
smaller. System studied in chemical thermodynamics (CTh) consist of several
{\em homogeneous macroscopic} phases cf. \cite{guggenheim67}. Evidently,
macroscopic statistics as in Chemistry cannot and should not be applied.
Taking this serious, fascinating perspectives open for statistical nuclear
fragmentation.
\end{abstract}
\maketitle
\section{introduction}Boltzmann developed statistical mechanics to explain
thermodynamics on a microscopic and mechanical basis. To that purpose, he
introduced the microcanonical ensemble of all realizations of a given
many-body system under the few macroscopic constraints of particle number
$N$, energy $E$ and its external volume $V$. The number of cells, with size
$(2\pi\hbar)^{3N}$, in the $6N$-dim phase space of such systems is
$W(E,N,V)$. Then he identified the thermodynamic entropy by the famous
formula which written on his tomb-stone:
\begin{equation} \fbox{\fbox{\vecbm{S=k*lnW}}}\label{boltzmann0}
\end{equation}
The physical meaning of this formula is: The entropy measures the amount of
our microscopic ignorance of the considered many-body system. If we would
have a complete knowledge, if we would know the positions and the momenta
of all particles at time $t$ (within quantum-mechanical uncertainty). $W$
would be a single cell in the $6N$-dim. phase space of size
$(2\pi\hbar)^{3N}$ and $S$ would be zero.

If our system is {\em homogeneous} (in a single phase) then the three
fundamental ensembles of statistical mechanics, the grand-canonical and the
canonical are equivalent to the microcanonical one in the thermodynamic
limit $N\Rightarrow\infty|_{\rho=N/V=const}$.

{\em Otherwise, however, they are not equivalent}. As only the above
equivalence of the ensembles guarantees the physical meaning of the
(canonical) ensembles, these have no physical justifications whenever the
equivalence is not given.

To apply statistical mechanics to such inhomogeneous systems like small
systems, or systems, even macroscopic, at phase separation, or very large
self-gravitating systems the following considerations must be taken
seriously.
\section{Macroscopic systems in Chemistry}
Microcanonical Boltzmann-Planck statistics is the basic of any
statistical mechanics, see e.g. Landau-Lifshitz \cite{landau96},
even Guggenheim defines statistical mechanics starting from the
microcanonical ensemble (\cite{guggenheim67}, chapter 2), see also
my paper http://arxiv.org/abs/cond-mat/0411408 v3.

Systems studied in chemical thermodynamics consist of several {\em
homogeneous macroscopic} phases $\alpha_1,\alpha_2,\cdots$
cf.\cite{guggenheim67}. Their mutual equilibrium must be explicitly
constructed from outside.

Each of these phases are assumed to be macroscopic (in the "thermodynamic
limit" ($N_\alpha\to\infty|_{\rho_\alpha=const}$)). There is no common
canonical ensemble for the entire system of the coexisting phases. Only the
canonical ensemble of {\em each} phase {\em separately} becomes equivalent
in the limit to its microcanonical counterpart.

The canonical partition sum of {\em each} phase $\alpha$ is defined as the
Laplace transform of the underlying  microcanonical sum of states
$W(E)_\alpha=e^{S_\alpha(E)}$ \cite{gross147,gross158}
\begin{equation}
Z_\alpha(T)= \int_0^\infty e^{S_\alpha(E)-E/T_\alpha} dE.
\end{equation}
The mean canonical energy is
\begin{eqnarray}
 <E_\alpha(T_\alpha)>&=&-\; \partial \ln Z_\alpha(T_\alpha)/\partial \beta_\alpha.\\
 \beta_\alpha&=&\frac{1}{T_\alpha}.\nonumber\\ \nonumber
\end{eqnarray}
In chemical situations proper the assumption of macroscopic
individual phases is of course acceptable. In the thermodynamic
limit ($N_\alpha\to\infty|_{\rho_\alpha=const}$) of a {\em
homogeneous} phase $\alpha$, the canonical energy
$<E_\alpha(T_\alpha)>$ becomes identical to the  microcanonical
energy $E_\alpha$ when the temperature is determined by
\begin{eqnarray}
T_\alpha^{-1}=\beta_\alpha&=&\left.\frac{\partial
S_\alpha(E,V_\alpha)}{\partial E}\right|_{E_\alpha}.\\ \nonumber
\end{eqnarray}
The relative width of the canonical energy is
\begin{equation}
\Delta
E_\alpha=\frac{\sqrt{<E_\alpha^2>-<E_\alpha>^2}}{<E_\alpha>}\propto
\frac{1}{\sqrt{N_\alpha}}.
\end{equation}
The heat capacity at constant volume is (care must be taken about
the constraints (!))
\begin{eqnarray}
C_\alpha|_{V_\alpha}&=&\frac{\partial
<E_\alpha(T_\alpha,V_\alpha)>}{\partial T_\alpha}\\ \nonumber\\
&=&\frac{<E_\alpha^2>-<E_\alpha>^2}{T_\alpha^2}\ge 0.\label{specheat}
\end{eqnarray}

Only in the thermodynamic limit ($N_\alpha\to\infty|_{\rho_\alpha=const}$)
does the energy uncertainty $\Delta E_\alpha\rightarrow 0$, and the
canonical and the microcanonical ensembles become equivalent. {\em I do not
know of any microscopic foundation of the canonical ensemble apart from the
limit.}

The positiveness of any canonical $C_V(T)$ or $C_P(T)$ is of
course the reason why the inhomogeneous system of several
coexisting phases ($\alpha_1 \& \alpha_2$) with an overall {\em
negative} heat capacity cannot be described by a {\em single
common} canonical distribution \cite{gross159,gross174}.
\section{Application to nuclear fragmentation}
Now, certainly, neither the phase of the whole multi-fragmented nucleus nor
the individual fragments themselves can be considered as macroscopic
homogeneous phases in the sense of chemical thermodynamics (CTh).
Consequently, (CTh) cannot and should not be applied to fragmenting nuclei
and the microcanonical description is ultimately demanded. This becomes
explicitly clear by the fact that the configurations of a multi-fragmented
nucleus have a {\em negative} heat capacity at constant volume $C_V$ and
also at constant pressure $C_P$ (if at all a pressure can be associated to
nuclear fragmentation \cite{gross174}).

However, there is a deep and fascinating aspect of {\em nuclear}
fragmentation:  First, in nuclear fragmentation we can measure the whole
statistical {\em distribution} of the ensemble event by event. Not only
their mean values are of physical interest. {\em Statistical Mechanics can
be explored from its first microscopic and mechanical principles in any
detail well away from the thermodynamic limit}. This give us an immediate
insight how statistical mechanics works. We can {\em measure} the entire
distribution also if they are not sharply Gaussian. No exotic non-extensive
statistics like Tsallis statistics \cite{tsallis88,tsallis99} is demanded
for equilibrized non-extensive systems. The "Sancta sanctorum of
statistical mechanics" i.e. the microcanonical Boltzmann-Planck statistics
covers all equilibrium systems \cite{gross203}.

Second, and this may be more important: For the first time phase
transitions to non-homogeneous phases can be studied where these phases are
within themselves composed out of several nuclei. This situation is very
much analogous to multi star systems like rotating double stars during
intermediate times, where nuclear burning prevents their final implosion.
The occurrence of negative heat capacities is an old well known peculiarity
of the statistics of self-gravitating systems \cite{thirring70}. Also these
cannot be described by a canonical ensemble. It was shown in
\cite{gross207,gross203} how the {\em microcanonical} phase space of these
self-gravitating systems has all the realistic configurations which are
observed, more details in my WEB-page http://www.hmi.de/people/gross/. Of
course, the question whether these systems are interim equilibrized or not
is not proven by this observation though it is rather likely.


\begin{thebibliography}{1}

\bibitem{guggenheim67}
E.A. Guggenheim.
\newblock {\em Thermodynamics, An Advanced Treatment for Chemists and
  Physicists}.
\newblock North-Holland Personal Library, Amsterdam, 1967.

\bibitem{landau96}
L.D. Landau and E.M. Lifshitz.
\newblock {\em Statistical Physics}.
\newblock Butterworth-Heinemann, Oxford, 1996.

\bibitem{gross147}
O.~Schapiro, D.H.E. Gross, and A.~Ecker.
\newblock Microcanonical Monte Carlo.
\newblock In P.J.-S~Shiue H.~Niederreiter, editor, {\em First International
  Conference on Monte Carlo and Quasi-Monte Carlo Methods in Scientific
  Computing}, volume 106, pages 346--353, Las Vegas, Nevada, 1995.

\bibitem{gross158}
D.H.E. Gross and M.E. Madjet.
\newblock Microcanonical vs. canonical thermodynamics.
\newblock {\em http://xxx.lanl.gov/abs/cond-mat/9611192}, 1996.

\bibitem{gross159}
D.H.E. Gross and M.E. Madjet.
\newblock Cluster fragmentation, a laboratory for thermodynamics and
  phase-transitions in particular.
\newblock In Abe, Arai, Lee, and Yabana, editors, {\em Proceedings of
  ''Similarities and Differences between Atomic Nuclei and Clusters''}, pages
  203--214, Tsukuba, Japan 97, 1997. The American Institute of Physics.

\bibitem{gross174}
D.H.E. Gross.
\newblock {\em Microcanonical thermodynamics: Phase transitions in ``Small''
  systems}, volume~66 of {\em Lecture Notes in Physics}.
\newblock World Scientific, Singapore, 2001.

\bibitem{tsallis88}
C.~Tsallis.
\newblock Possible generalization of Boltzmann-Gibbs statistics.
\newblock {\em J.Stat.Phys}, 52:479, 1988.

\bibitem{tsallis99}
C.~Tsallis.
\newblock Nonextensive statistics: Theoretical, experimental and computational
  evidences and connections.
\newblock {\em Braz.Journ.Physics}, 29:1, 1999.

\bibitem{gross203}
D.H.E. Gross.
\newblock Classical equilibrium thermostatistics, "Sancta sanctorum of
  statistical mechanics", from nuclei to stars.
\newblock {\em Physica A}, 340/1-3:76--84, cond--mat/0311418, (2004).

\bibitem{thirring70}
W.~Thirring.
\newblock Systems with negative specific heat.
\newblock {\em Z. f. Phys.}, 235:339--352, 1970.

\bibitem{gross207}
D.H.E. Gross.
\newblock A new thermodynamics from nuclei to stars.
\newblock {\em Entropy}, 6:158--179, (2004).

\end{thebibliography}

\end{document}